\documentclass[aps,pra,10pt,twocolumn,superscriptaddress]{revtex4}

\usepackage{amsmath}
\usepackage{amssymb}
\usepackage{bbm}
\usepackage{framed}
\usepackage{color}
\usepackage[colorlinks = true, urlcolor = blue]{hyperref}
\usepackage{epstopdf}
\usepackage{dsfont}
\usepackage{amssymb}
\usepackage{amsmath}
\usepackage{amsthm}
\usepackage{wrapfig}
\usepackage{relsize}
\usepackage{bm}
\usepackage[latin1]{inputenc}
\usepackage{enumitem}
\usepackage{natbib}
\usepackage{graphicx}
\usepackage{physics}
\usepackage{xcolor}
\usepackage{hyperref}

\begin{document}

	\title{Time-Polynomial Lieb-Robinson bounds for finite-range spin-network models}

	\author{Stefano Chessa}
	\email{stefano.chessa@sns.it}
	\author{Vittorio Giovannetti}
	\affiliation{NEST, Scuola Normale Superiore and Istituto Nanoscienze-CNR, I-56126 Pisa, Italy}
	\date{\today}

\begin{abstract}
The Lieb-Robinson bound sets
a theoretical upper limit on the speed at which information can propagate in non-relativistic quantum 
spin networks. In its original version, it results in an exponentially exploding function of the evolution time,
which is partially mitigated by an exponentially decreasing term that instead depends upon the
distance covered by the signal (the ratio between the two exponents effectively defining an upper bound on the propagation speed). In the present paper, by properly accounting for the free parameters of the model,
we show how to turn this construction into a stronger inequality where the upper limit only scales polynomially with respect to the evolution time. Our analysis applies to any chosen topology of the network, as long as the range of the associated interaction is  explicitly finite. 
For the special case of linear spin networks we present also an alternative derivation based on 
a perturbative expansion approach which improves the previous inequality. In the same context we also
establish a lower bound to the speed of the information spread which yields a non trivial result at least in the limit of small propagation times. 
\end{abstract}

\maketitle

\section{Introduction} \label{sec.Intro}

When dealing with communication activities, information transfer speed is one of the most relevant parameters in order to characterise the communication line performances. This statement applies both to Quantum Communication, obviously, and Quantum Computation, where the effective ability to carry information, for instance from a gate to another one, can determine the number of calculations executable per unit of time. It appears therefore to be useful being able to estimate such speed or, whenever not possible, bound it with an upper value. In the context of communication via quantum spin networks~\cite{BOSE1}  a result of this kind can be obtained exploiting the so called {Lieb-Robinson (L-R) bound} \cite{LR,REVIEW}: defining a suitable correlation function  involving two local spatially separated operators $\hat{A}$ and $\hat{B}$, a maximum group velocity for correlations and consequently for signals can be extrapolated. In more recent years this bound has been generalised and applied to attain results in a wider set of circumstances. Specifically, among others, stick out proofs for the {Lieb-Schultz-Mattis theorem in higher dimensions} \cite{LSM Theo}, for the {exponential clustering theorem} \cite{Clust Theo},
to link spectral gap and exponential decay of correlations for short-range interacting systems
\cite{exponential1},
 for the {existence of the dynamics for interactions with polynomial decay} \cite{ExistDynam}, for  {area law in 1-D systems} \cite{AreaLaw}, for the {stability of topological quantum order}~\cite{TopQOrder}, for information and entanglement spreading~\cite{BRAV,SUPER,EISERT,PRA}, for black holes physics and information scrambling~\cite{Scram, Scram1}.
  Bounds on correlation spreading, remaining in the framework set by L-R bounds, have been then generalized to different scenarios such as, for instance, long-range interactions \cite{LongRange, LongRange1, LongRange2, LongRange3, LongRange4}, disordered systems \cite{Burrell,Burrell2}, finite temperature \cite{FinTemp,FinTemp1,FinTemp2}. 
After the original work by Lieb and Robinson the typical shape found to describe the bound has been the exponentially growing in time $t$ and depressed with the spatial distance between the supports of the two operators $d(A,B)$, namely:
\begin{eqnarray}\label{ORIGIM} 
\|  [\hat{A}(t),\hat{B}] \|  \lesssim   \, e^{v|t|} \;  f( d(A,B))  \;,\end{eqnarray} 
with $v$  positive constant, and $f(\cdot)$ being a suitable decreasing function,  both depending upon the interaction considered, the size of the supports of $\hat{A}$ and $\hat{B}$ and the dimensions of the system~\cite{LSM Theo,Clust Theo,ExistDynam,exponential1}. 
More recently instances have been proposed~\cite{FinTemp2,Them} in which such behaviour can be improved to a polynomial one 
\begin{eqnarray} \|  [\hat{A}(t),\hat{B}] \|  \lesssim \left(\frac{t}{d(A,B)}\right)^{d(A,B)}\;, \label{LEQSIMM} \end{eqnarray} at least for Hamiltonian couplings which have an explicitly finite range, and for short enough times.  Aim of the present work is to set these  results on a firm ground
providing 
 an alternative derivation of  the polynomial version~(\ref{LEQSIMM}) of the  L-R inequality which, as long as the range of the interactions involved is  finite, holds true
for arbitrary topology of the spin network and which does not suffer from the
short time limitations that instead affects previous approaches. 
Our analysis yields a simple way to estimate the maximum speed 
at which signals can propagate along the network. 
In the second part of the manuscript we focus instead on the special case of single sites located at the extremal points of a 1-D linear spin chain model.
In this context we give an alternative derivation of the $t$-polynomial L-R bound and discuss how the same
technique can also be used to provide a lower bound on $\|  [\hat{A}(t),\hat{B}] \|$, which at least for
small $t$ is non trivial.

The manuscript is organized as follows.
We start in Sec.~\ref{sec1} presenting the model and recalling the 
original version of the L-R bound. The main result of the paper is hence presented in 
Sec.~\ref{sec1new} where by using simple analytical argument we derive our 
$t$-polynomial version of the L-R inequality. In Sec.~\ref{SEC:PERT} we present instead
the perturbative expansion approach for 1-D linear spin chain models. In Sec.~\ref{Sec:Simulation} we test results achieved in previous sections by comparing them to the numerical simulation of a spin chain.
Conclusions are presented finally in Sec.~\ref{Sec:conc}.

\section{The model and some preliminary observations} \label{sec1} 

Adopting the usual framework for the derivation of the L-B bound~\cite{Clust Theo}
let us consider a network $\cal N$ of quantum systems (spins) distributed on a graph $\mathbb{G}:=(V,E)$ characterized by a set of
 vertices $V$ and by a set $E$ of edges. The model is  equipped with a metric  $d(x,y)$ defined as the shortest path (least number of edges) connecting $x,y \in  V$ ($d(x,y)$ being set equal to infinity in the absence of a connecting path), which induces a measure for 
the diameter $D(X)$ of a given subset $X\subset V$, and a  distance $d(X,Y)$ among
the elements  $X,Y \subset V$,  

\begin{eqnarray} 
   D(X)&:=&\max\limits_{x,y} \min \{ d(x,y) | x,y\in X\}\;, \nonumber \\ 
   d(X,Y) &:=& \min \{ d(x,y) | x \in X ,y\in Y\}\;. \end{eqnarray} 

Indicating  with ${\cal H}_x$ the Hilbert space associated with spin that occupies the vertex $x$ of the graph, 
 the  Hamiltonian of ${\cal N}$ can be formally written as 
\begin{eqnarray} \label{HAMILT} 
\hat{H} : = \sum_{X\subset V} 
\hat{H}_X\;, 
\end{eqnarray} 
where the summation runs over the  subsets $X$ of $V$ with $\hat{H}_X$ being a self-adjoint operator that  is local on the Hilbert space  
 ${\cal H}_X:=  \otimes_{x\in X} {\cal H}_x$ , i.e. it acts 
 non-trivially on  the spins of $X$ while being the identity everywhere else. 
       Consider then two
 subsets $A,B\subset V$ which are disjoint,  $d(A,B)>0$. Any two operators
   $\hat{A}:= \hat{A}_{A}$ and $\hat{B}:=\hat{B}_{B}$ that are local  on such subsets clearly commute, i.e. 
         $[\hat{A},\hat{B}]=0$. Yet as we let the system evolve under the action of the Hamiltonian $\hat{H}$,  this condition will not necessarily hold 
         due to the building up of correlations along the graph.
         More precisely, given $\hat{U}(t):= e^{-i \hat{H}t}$ the unitary evolution induced by (\ref{HAMILT}), and indicating with 
         \begin{eqnarray} 
           \hat{A}(t) := \hat{U}^\dag(t) \hat{A} \hat{U}(t)\;,
         \end{eqnarray} 
           the evolved counterpart of $\hat{A}$ in the Heisenberg representation, we expect the 
           commutator $[\hat{A}(t), \hat{B}]$ to become explicitly non-zero for large enough $t$, the faster this happens, the strongest being the correlations
           that are dynamically induced by $\hat{H}$ (hereafter we set $\hbar =1$ for simplicity).          The Lieb-Robinson bound puts a limit on such behaviour that applies for 
           all $\hat{H}$ which are characterized by  couplings that have a  finite range character (at least approximately). Specifically, 
indicating with  $|X|$ the total number of sites in  the domain $X\subset V$,  and with 
\begin{eqnarray} M_X:= \max_{x \in X} \mbox{dim}[{\cal H}_x]\;, 
\end{eqnarray}
 
the  maximum value of its spins Hilbert space dimension, we say that $\hat{H}$ is well behaved in terms of long range interactions, if there exists a positive constant $\lambda$ such that the functional 
\begin{equation} \label{DEFPHILambda} 
\| \hat{H}\|_\lambda :=\sup_{x \in V}  \sum\limits_{X\ni x}  \left|X\right| M_X^{2\left|X\right|} e^{\lambda D(X)} \; \|\hat{H}_X\|  \, ,
\end{equation}
is finite.  In this expression the 
 symbol 
 \begin{eqnarray} \| \hat{\Theta} \|: = \max_{|\psi\rangle} \| \hat{\Theta} |\psi\rangle\|\;,\end{eqnarray}  represents the standard operator norm, while the summation runs over all the subset $X \subset V$ that contains $x$ as an element. Variant versions~\cite{Bratteli, Clust Theo,exponential1} or generalizations~\cite{REVIEW,NACHTER1} of Eq.~(\ref{DEFPHILambda}) can be found in the literature%see e.g.  \cite{Bratteli, Clust Theo,exponential1,REVIEW}
, however as they express the same behaviour and substantially differ only by constants, in the following we shall gloss over these differences.
 The L-R bound can now be expressed in the form of the following inequality~\cite{Clust Theo} 
\begin{equation}\label{eq: lambda bound}
\| [\hat{A}(t),\hat{B}] \| 
\leq 
2 |A| | B| \| \hat{A}\| \| \hat{B}\| ( e^{2 |t| \| \hat{H} \|_\lambda}-1) e^{ -\lambda\, d(A,B)}\;,
\end{equation}
which holds non trivially for well behaved Hamiltonian $\hat{H}$ admitting  finite values of the quantity $\| \hat{H} \|_\lambda$. 
It is worth stressing that Eq.~(\ref{eq: lambda bound}) is valid irrespectively from the initial state of
the network and that, due to the dependence upon $|t|$ on the r.h.s. term, exactly the same bound  can be derived for $\| [\hat{A},\hat{B}(t)] \|$, obtained by exchanging the roles of
$\hat{A}$ and $\hat{B}$. Finally we also point out that in many cases of physical interest the pre-factor $|A| | B|$ on the r.h.s. can be simplified: for instance
it can be omitted for one-dimensional models, while for nearest neighbor interactions one can replace this by the smaller of the boundary sizes of $\hat{A}$ and $\hat{B}$ supports~\cite{NACHTER1}.

For models characterized by  interactions which are explicitly not finite, refinements of Eq.~(\ref{eq: lambda bound}) have been obtained under special constraints on the decaying of the 
long-range Hamiltonian coupling contributions~\cite{exponential1,Clust Theo}.  For instance assuming that there exist  (finite) positive  quantities $s_1$ and $\mu_1$ ($s_1$ being independent from total number of sites of the graph  $\mathbb{G}$), such that 
\begin{equation} \label{DEFPHILambdaNEW} 
\sup_{x \in V}  \sum\limits_{X\ni x}  \left|X\right|  \|\hat{H}_X\| [1+D(X)]^{\mu_1} \leq s_1   \, ,
\end{equation}
one gets 
\begin{equation}\label{eq: lambda boundnew1}
\| [\hat{A}(t),\hat{B}] \| 
\leq 
C_1 |A| | B| \| \hat{A}\| \| \hat{B}\| \frac{ e^{v_1 |t|} -1}{ (1+ d(A,B))^{\mu_1}} \;,
\end{equation}
with $C_1$ and $v_1$ positive quantities that only depend upon the metric of the network and on the Hamiltonian. 
On the contrary if there exist  (finite) positive quantities $\mu_2$ and $s_2$ (the latter being again  independent from total number of sites of $\mathbb{G}$),  such that 
\begin{equation} \label{DEFPHILambdaNEW2} 
\sup_{x \in V}  \sum\limits_{X\ni x}  \left|X\right|  \|\hat{H}_X\|  e^{\mu_2 D(X)}  \leq s_2   \, ,
\end{equation}
we get instead 
\begin{equation}\label{eq: lambda boundnew2}
\| [\hat{A}(t),\hat{B}] \| 
\leq 
C_2 |A| | B| \| \hat{A}\| \| \hat{B}\| ( e^{v_2 |t|} -1)  e^{-\mu_2 d(A,B)}  \;, 
\end{equation}
where once more  $C_2$ and $v_2$ are positive quantities that only depend upon the metric of the network and on the Hamiltonian. 
The common trait of these results is the fact that their associated upper bounds maintain 
the exponential dependence  with respect to the transferring $t$ enlightened in Eq.~(\ref{ORIGIM}). 

\section{Casting the Lieb-Robinson bound into a $t$-polynomial form for (explicitly) finite range couplings} \label{sec1new} 

 The  inequality  (\ref{eq: lambda bound}) is the starting point of our analysis: it is indicative of the fact that the model admits a finite speed $v\simeq 2  \| \hat{H} \|_\lambda /
 \lambda$ at which correlations can spread out  in the spin network. 
  As $|t|$ increases, however,  the bound becomes
 less and less informative due to the exponential dependence of the r.h.s.: in particular it  becomes irrelevant as soon as the multiplicative factor 
 of $\| \hat{A}\| \| \hat{B}\|$ 
 gets larger than $2$. In this limit in fact Eq.~(\ref{eq: lambda bound}) is trivially subsided by the inequality 
 \begin{eqnarray} \| [\hat{A}(t),\hat{B}] \| \leq 2 \| \hat{A}(t)\| \| \hat{B}\|= \label{TRIVIAL} 
 2 \| \hat{A}\| \| \hat{B}\|\;, \end{eqnarray}  that follows by simple
 algebraic considerations. 
One way to strengthen the conclusions one can draw from (\ref{eq: lambda bound}) 
is  to consider $\lambda$ as a free parameter and to optimize with respect to all the values it can assume.
As the functional dependence of $\| \hat{H} \|_\lambda$ upon $\lambda$ is strongly influenced by  the specific properties
of the 
spin model, we restrict the analysis to the special (yet realistic and interesting) scenario
of 
 Hamiltonians $\hat{H}$~(\ref{HAMILT})  which
are strictly short-ranged. Accordingly we now impose  $\hat{H}_X=0$ to all the
subsets $X\subset V$ which have a diameter $D(X)$ that is larger than  a fixed finite value $\bar{D}$, i.e. 
 \begin{eqnarray} 
\quad D(X) > \bar{D} \qquad  \Longrightarrow \qquad \hat{H}_X = 0  \;, 
 \end{eqnarray} 
which is clearly  more stringent than 
 both those presented in  Eqs.~(\ref{DEFPHILambdaNEW}) and~(\ref{DEFPHILambdaNEW2}). 
 Under this condition $\hat{H}$ is well behaved for all $\lambda \geq 0$ and one can write  
 \begin{equation}\label{eq: Max diameter}
 \|\hat{H} \|_{\lambda}\leq \zeta \, e^{\lambda\bar{D}}, \qquad \forall \lambda \geq 0\;, 
\end{equation} 
with  $\zeta$ being a finite positive constant that for sufficiently regular graphs does not scale 
with the total number of spins of the system.
For instance for regular arrays of 
 first-neighbours-coupled spins  we get $\zeta = 2  C M^4 \| \hat{h}\|$,
where $C$ is the maximum coordination number of the graph (i.e. the number of
edges associated with a given site), 
\begin{eqnarray} 
\| \hat{h}\| : = \sup_{X\subset V} \| \hat{H}_X\|\;,\end{eqnarray}  is the maximum  strength of the interactions, and  where  $M:=  \max_{x \in V} \mbox{dim}[{\cal H}_x]$ is the maximum dimension of the local spins Hilbert space
of the model. 
More generally for graphs $\mathbb{G}$ characterized by finite values of $C$ it is easy to show that $\zeta$ can
 not be greater than $C^{\bar D} M^{C^{\bar D}} \| \hat{h} \|$.

\begin{figure}[t!]
  \includegraphics[width=\linewidth]{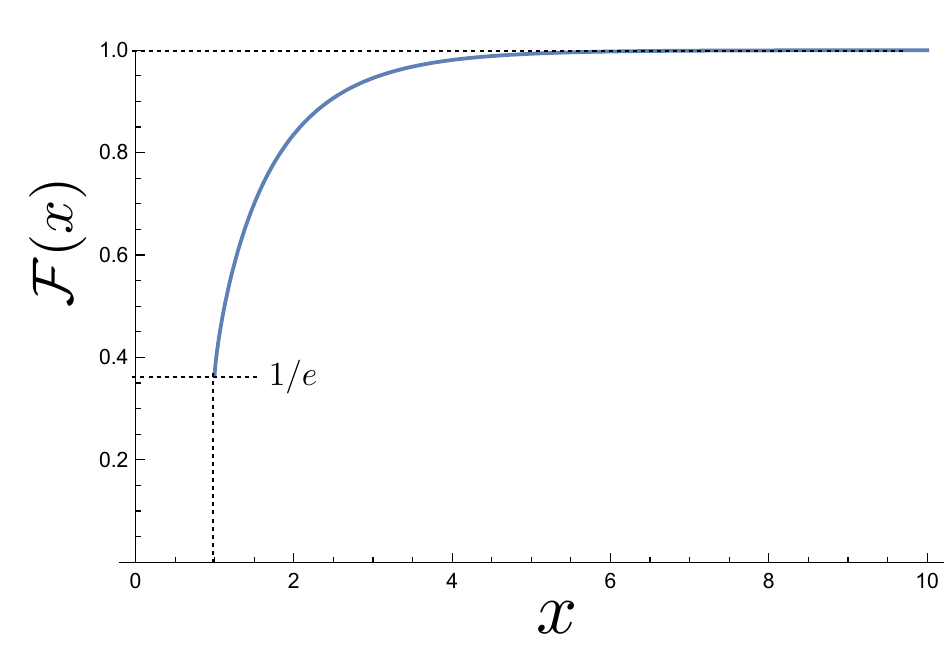}
  \caption{Plot of the function ${\cal F}(x)$ entering into the derivation of Eq.~(\ref{eq:MinNachBound}): 
for  $x=\tfrac{d(A,B)}{\bar{D}}\geq 1$ it is monotonically increasing reaching the value 
$1/e\simeq 0.37$ for $x=1$ and quickly approaching the asymptotic value $1$ for large enough $x$.}
  \label{fig:plotC}
\end{figure}

 Using  (\ref{eq: Max diameter}) we can now turn~(\ref{eq: lambda bound}) into a more 
treatable expression
\begin{equation}
\| [\hat{A}(t),\hat{B}] \|   \leq  
2 |A| |B|  \| \hat{A}\| \| \hat{B}\|   
(e^{2 |t| \zeta e^{\lambda \bar{D}}}  -1) e^{ -\lambda\, d(A,B)}\;, 
\label{eq: lambda boundnew complete}
\end{equation}
whose r.h.s. can now be explicitly minimized in terms of $\lambda$ for any fixed $t$ and 
 $d(A,B)$. As shown  in Sec.~\ref{APPimp} the final result is given by  
 \begin{eqnarray} 
 \nonumber 
\| [\hat{A}(t),\hat{B}] \|  &\leq& 2 |A| |B|  \| \hat{A}\| \| \hat{B}\|  \left(\frac{2\, e\,\zeta\,\bar{D}\,|t|}{d(A,B)}\right)^{\tfrac{d(A,B)}{\bar{D}}}  \!\!\!\!\!\! \!\! {\cal F}(\tfrac{d(A,B)}{\bar{D}}) \\
 &\leq& 2|A| |B|   \| \hat{A}\| \| \hat{B}\|  \left(\frac{2\, e\,\zeta\,\bar{D}\,|t|}{d(A,B)}\right)^{\tfrac{d(A,B)}{\bar{D}}}  \!\!\!\!\!\!\;, 
 \label{eq:MinNachBound}
\end{eqnarray}
where in the second inequality we used the fact that the function ${\cal F}(x)$ defined in the Eq.~(\ref{DEFC}) below and  plotted in  
Fig.~\ref{fig:plotC} is monotonically increasing and bounded from above by its asymptotic value $1$.
At variance with Eq.~(\ref{eq: lambda bound}), the inequality~(\ref{eq:MinNachBound}) contains only terms which are explicit functions of the spin network parameters.   
Furthermore the new bound is polynomial in $t$ with a scaling that is definitely better 
than the linear behaviour one could infer from the Taylor expansion of the r.h.s. of Eq.~(\ref{eq: lambda bound}). Looking at the spatial component of (\ref{eq:MinNachBound}) we notice that correlations still decrease with distance as well as in bounds (\ref{eq: lambda bound}), (\ref{eq: lambda boundnew1}) and (\ref{eq: lambda boundnew2}) but with a scaling $(1/x)^x=e^{-x \log x}$ that is more than exponentially depressed.
Also, fixing a (positive) target threshold value $R_*<1$ for the ratio 
\begin{eqnarray} 
R(t):=\| [\hat{A}(t),\hat{B}] \| /(2 |A| |B|  \| \hat{A}\| \| \hat{B}\| )\;, \end{eqnarray}  equation~(\ref{eq:MinNachBound}) predicts that it will be reached not before a time interval
\begin{eqnarray}
t_*  = \frac{d(A,B) R_*^{\bar{D}/d(A,B)}}{2 e \zeta \bar{D}}  \;, \label{elapsingtime} 
\end{eqnarray} 
has elapsed from the beginning of the dynamical evolution.
Exploiting the fact that $\lim_{z\rightarrow \infty} R_*^{1/z}=1$, in the asymptotic limit of very distant sites
 (i.e. $d(A,B)\gg \bar{D}$), this can be simplified to 
 \begin{eqnarray}
t_*  \simeq \frac{d(A,B) }{2 e \zeta \bar{D}}  \;, \label{elapsingtime} 
\end{eqnarray}
that is independent from the actual value of the target $R_*\neq 0$, leading us to identify the quantity 
\begin{eqnarray} \label{SPEED} 
v_{\max} := 2 e \zeta \bar{D}\;,
\end{eqnarray} 
as an upper bound for the maximum speed allowed for the propagation  of signals in the system.

\subsection{Explicit derivation of Eq.~(\ref{eq:MinNachBound}) }\label{APPimp}

We start by noticing that by neglecting the negative contribution $-e^{ -\lambda\, d(A,B)}$, we can bound the r.h.s. Eq.~(\ref{eq: lambda boundnew complete}) by a
 form which is much easier to handle, i.e.  
\begin{eqnarray}
\| [\hat{A}(t),\hat{B}] \|    
\leq 2|A| |B|  \| \hat{A}\| \| \hat{B}\|  e^{2 |t| \zeta e^{\lambda \bar{D}}   -\lambda\, d(A,B)}.\label{eq: lambda boundnew}
\end{eqnarray}
One can observe that 
for $t > d(A,B)/ (2 \zeta \bar{D})$ the approach 
 yields an   inequality that
is always less stringent than (\ref{TRIVIAL}). On the contrary for  $|t| \leq d(A,B)/ (2 \zeta \bar{D})$, 
 imposing  the stationary condition on the exponent term, i.e.  
$\partial_{\lambda}(e^{2\,\zeta\, e^{\lambda \overline{D}(X)}|t| -\lambda\, d(A,B)})=0$, 
 we found that for  the 
 optimal value for $\lambda$ is provided by 
\begin{equation}
\lambda_{\rm opt}:=\frac{1}{\bar{D}}\ln(\frac{d(A,B)}{2\,|t|\,\zeta\,\bar{D}}),
\end{equation}
which replaced 
in Eq.~(\ref{eq: lambda boundnew}) yields  directly (\ref{eq:MinNachBound}).
More generally, we can avoid to pass through Eq.~(\ref{eq: lambda boundnew}) 
 by 
looking for minima of the r.h.s. of Eq.~(\ref{eq: lambda bound}) obtaining the first inequality given in Eq.~(\ref{eq:MinNachBound}), i.e. 
\begin{equation} \label{eq:MinNachBound11}
\| [\hat{A}(t),\hat{B}] \|  \leq2|A| |B|  \| \hat{A}\| \| \hat{B}\|  \left(\tfrac{2\, e\,\zeta\,\bar{D}\,|t|}{d(A,B)}\right)^{\tfrac{d(A,B)}{\bar{D}}} \!\!\!
{\cal F}(\tfrac{d(A,B)}{\bar{D}})  \;.
\end{equation}

For this purpose we consider a parametrization of the coefficient 
  $\lambda$ in terms of the positive variable $z$ as indicated here
\begin{equation}
\lambda:=\frac{1}{\bar{D}}\ln(\frac{ z  d(A,B)}{2\,|t|\,\zeta\,\bar{D}}).
\end{equation} 
With this choice the quantity we are interested in becomes 
\begin{eqnarray} 
\label{FFDD1} 
&&2 |A| |B|  \| \hat{A}\| \| \hat{B}\|   (e^{2 |t| \zeta e^{\lambda \bar{D}}}  -1) e^{ -\lambda\, d(A,B)} \\ \nonumber 
&&\qquad \qquad\qquad= 2 |A|  |B|  \| \hat{A}\| \| \hat{B}\|   \left(\tfrac{2 e t \zeta }{x} \right)^{x} f_x
(z) \;, 
\end{eqnarray} 
where in the r.h.s. term for easy of notation we introduced  $x=d(A,B)/\bar{D}$ and the function
\begin{eqnarray} \label{DEFF} 
f_x(z):= \frac{e^{x z} -1}{z^x e^x} \;.\end{eqnarray} 
For fixed value of $x\geq 1$ the minimum of the Eq.~(\ref{DEFF}) is attained for $z=z_{\rm opt}$ fulfilling the constraint  
\begin{eqnarray} 
x = - \frac{\ln(1-z_{\rm opt})}{z_{\rm opt}} \;.
\end{eqnarray}  By formally inverting this expression and by inserting it into Eq.~(\ref{FFDD1}) 
we   hence get (\ref{eq:MinNachBound11}) 
with
\begin{eqnarray} \label{DEFC} 
{\cal F}(x):= 
\frac{z_{\rm opt}(x)}{1-z_{\rm opt}(x)} \left(\frac{1}{e z_{\rm opt}(x)}\right)^x \;, 
\end{eqnarray} 
being the monotonically increasing function reported in Fig.~\ref{fig:plotC}.

\section{Perturbative expansion approach}\label{SEC:PERT} 

An alternative derivation of a $t$-polynomial bound similar to the one reported in Eq.~(\ref{eq:MinNachBound}) can be obtained by adopting 
 a perturbative expansion  of  the unitary evolution of the operator $\hat{A}(t)$ that allows one to express the commutator
 $[ \hat{A}(t), \hat{B}]$ as a sum over a collections of ``paths'' connecting the locations $A$ and $B$,  see e.g. Eq.~(\ref{summation}) below. 
This derivation is somehow analogous to the one used in Refs.~\cite{FinTemp2,Them}. 
Yet in these papers the number of relevant terms entering in the calculation of  the norm of $[ \hat{A}(t), \hat{B}]$
could be underestimated by just considering those paths which are obtained 
by concatenating 
adjacent contributions and resulting in corrections that are negligible only  for small times $t$.
 In what follows we shall overcome these limitations by focusing on the special case of linear spin chains which allows for a proper account of the relevant paths. Finally we shall  see how it is possible to exploit the perturbative expansion approach 
to also derive a lower bound for $\| [ \hat{A}(t), \hat{B}]\|$.

 While in principle the perturbative expansion approach can be adopted to discuss arbitrary topologies of the network, in order to get a closed formula for the final expression we shall restrict the analysis to the case of two single sites  (i.e. $|A|=|B|=1$)  located at the end of a 
$N$-long, 1-D spin chain with next-neighbour interactions (i.e. $d=N-1$).  Accordingly we shall write the Hamiltonian 
(\ref{HAMILT}) as 
\begin{eqnarray}  \label{1DMODEL} 
\hat{H}:=\sum\limits_{i=1}^{N-1} \hat{h}_i\;,
\end{eqnarray} 
with $\hat{h}_i$ operators acting non trivially only on the $i$-th and $(i+1)$-th spins, hence fulfilling the condition 
\begin{eqnarray} \label{COMM} 
[\hat{h}_i,\hat{h}_j]=0 \;, \qquad \forall |i-j|>1\;.
\end{eqnarray} 
%\textcolor{magenta}{To simplify the derivation we shall also assume that $\hat{A}$ and $\hat{B}$ operate locally on two individual spins at the ends of the chain
%(i.e. $|A|=|B|=1$)  separated by $d\geq 1$ lattice steps, s.t. in this case $N=d$.}

 \subsection{Upper bound} \label{Sec:UPPER}

  Adopting the Baker-Campbell-Hausdorff  formula we  write 
\begin{equation}\label{eq: CBH expansion}
  [\hat{A}(t),\hat{B}]  = [\hat{A},\hat{B}] + \sum\limits_{k=1}^\infty \frac{(it)^k}{k!} \left[[\hat{H},\hat{A}]_k,\hat{B} \right]\;,
\end{equation}
where for $k\geq 1$,  
\begin{eqnarray} [\hat{H},\hat{A}]_k:= [\overbrace{\hat{H},[\hat{H},[\cdots, [\hat{H}, [\hat{H}}^{k\mbox{ times}},\hat{A}]]\cdots]]\;,
\end{eqnarray}  indicates the $k$-th order, nested commutator between $\hat{H}$ and $\hat{A}$. 
Exploiting the structural properties of Eqs.~(\ref{1DMODEL})  and (\ref{COMM}) 
 it is easy to check that the only terms which may give us a non-zero contribution to the
r.h.s. of 
Eq.~(\ref{eq: CBH expansion})  are those with $k\geq d$. 
Accordingly we get 
\begin{equation}\label{eq: CBH expansion11}
  [\hat{A}(t),\hat{B}]  =  \sum\limits_{k=d}^\infty \frac{(it)^k}{k!} \left[[\hat{H},\hat{A}]_k,\hat{B} \right]\;,
 \end{equation} 
which leads to 
\begin{eqnarray}
\|  [\hat{A}(t),\hat{B}] \|  \leq     \sum\limits_{k=d}^\infty \frac{|t|^k}{k!} \| [[\hat{H},\hat{A}]_k,\hat{B} ] \| \;, \label{eq: Corr func all terms}
\end{eqnarray}
 via sub-additivity of the norm.
To proceed further we observe that 
  \begin{eqnarray} 
\| [[\hat{H},\hat{A}]_k,\hat{B}]\|\leq 2\|  \hat{A} \|  \|  \hat{B} \|  \| 2 \hat{H} \|^k \;,
\end{eqnarray} 
which for sufficiently small times $t$ yields
\begin{eqnarray}
\| [\hat{A}(t),\hat{B}] \|  &\simeq& \frac{|t|^{d}}{d!}\|  [[\hat{H},\hat{A}]_d,\hat{B}]\| 
\nonumber \\
&\leq& 2\|  \hat{A} \|  \|  \hat{B} \|  \frac{\left( 2\|  
\hat{H} \|  |t| \right)^{d}}{d!} \nonumber  \\ 
\label{eq: Poly bound1}
  &\leq& \frac{ 2\|  \hat{A} \|  \|  \hat{B} \| }{\sqrt{2\pi d}} \left( \frac{2\,e\|  \hat{H} \| \, |t| }{d} \right)^{d},
\end{eqnarray}
where in the last passage we adopted the lower bound on $d!$ that follows from   the  Stirling's inequalities 
\begin{eqnarray} 
(d/e)^d \sqrt{e^2 d} \geq 
d!\geq (d/e)^d \sqrt{2\pi d}\;,
\label{STIRLING} \end{eqnarray} 
Equation~(\ref{eq: Poly bound1})  exhibits a 
 polynomial behaviour similar to the one observed in Eq.~(\ref{eq:MinNachBound}) 
 (notice that if instead of next-neighbour we had next-$\bar{D}$-neighbours interaction the first not null order will be the $\lceil \frac{d}{\bar{D}} \rceil$-th one and accordingly, assuming ${d}/{\bar{D}}$ to be integer, the above derivation will still hold with $d$ replaced by $d/\bar{D}$). 
  Yet the derivation reported above
 suffers from two main limitations: first of all it only holds for sufficiently small $t$ due to the fact that we have neglected all the terms 
 of (\ref{eq: Corr func all terms}) but the first one; second  the r.h.s of  Eq.~(\ref{eq: Poly bound1}) has a direct dependence on the total size $N$
   of the system carried by $\|  \hat{H} \|$, i.e. on the distance $d$ connecting the two sites.
Both these problems can be avoided by carefully considering
 each ``nested'' commutator $[[\hat{H},\hat{A}]_k,\hat{B}]$ entering
~(\ref{eq: Corr func all terms}). Indeed given the structure of the 
 Hamiltonian and the linearity of commutators, it follows that we can write 
 \begin{eqnarray} \label{summation} 
 [[\hat{H},\hat{A}]_k,\hat{B}] =\sum_{i_1,i_2,\cdots, i_k=1}^{N-1} [\hat{C}^{(k)}_{i_1,i_2 ,\cdots ,i_k} (\hat{A}) ,\hat{B} ]\;,\end{eqnarray} 
 where for 
  $i_1 ,i_2 ,\cdots ,i_k\in\{1,2,\cdots,N-1\}$ we have 
 \begin{eqnarray}\label{Knested} 
\hat{C}^{(k)}_{i_1,i_2 ,\cdots ,i_k} (\hat{A}) :=[\hat{h}_{i_k} , [\hat{h}_{i_{k-1}} , \cdots ,[\hat{h}_{i_2}, [\hat{h}_{i_1} ,\hat{A}]]\cdots]]\;.
\end{eqnarray} 
Now taking into account the commutation rule~(\ref{COMM}) and of the fact 
that $\hat{A}$ and $\hat{B}$ are located at the two opposite ends of the chain, it turns out that only a limited number  
\begin{eqnarray} n_k \leq \binom{k}{d} d^{k-d} = \frac{k! \; d^{k-d}}{d! (k-d)!} \label{BOUNDONNK} 
\;,\end{eqnarray}
 of the $N^k$ terms entering (\ref{summation}) 
 will have a chance of being  non zero. For the sake of readability we postpose the explicit derivation of this inequality (as well as the comment on alternative approaches presented in Refs.~\cite{FinTemp2,Them})
   in Sec.~\ref{sec. Counting comm}: here instead we observe that 
using 
 \begin{eqnarray} \| [\hat{C}^{(k)}_{i_1,i_2 ,\cdots ,i_k} (\hat{A}) ,\hat{B} ] \|\leq 2 \| \hat{A}\| \| \hat{B}\| (2 \| \hat{h}\|)^k \;, 
 \end{eqnarray} 
 where now $\| \hat{h}\| := \max\limits_i\|  \hat{h}_i \|  $, it allows us to  transform Eq.~(\ref{eq: Corr func all terms}) into 
\begin{eqnarray}
\|  [\hat{A}(t),\hat{B}] \|  &\leq& 2\|  \hat{A} \|  \|  \hat{B} \|  \sum_{k=d}^\infty n_k\frac{(2 |t| \|  \hat{h} \| )^k}{k!}
\nonumber \\ 
& \leq &2\|  \hat{A} \|  \|  \hat{B} \| \frac{\left( 2 |t|\|  \hat{h} \| \right)^{d}}{d!}\sum_{k=0}^\infty \frac{\left( 2
|t| \|  \hat{h} \|  d\right)^k}{k!}\nonumber \\ 
&=&2\|  \hat{A} \|  \|  \hat{B} \| \frac{\left(2|t|\|  \hat{h} \| \right)^{d}}{d!}e^{2|t|\|  \hat{h} \| d} \;, \nonumber \end{eqnarray}
which presents a scaling that closely resemble to one obtained in Ref.~\cite{CRAMER} for finite-range quadratic Hamiltonians for  harmonic systems on a lattice. 
Invoking hence  the lower bound for $d!$ that follows from  (\ref{STIRLING}) we finally get 
\begin{equation}  \label{UUPP} 
\|  [\hat{A}(t),\hat{B}] \|    \leq \frac{2\|  \hat{A} \|  \|  \hat{B} \|}{  \sqrt{2 \pi d} }   \left( \frac{2\,e\| \hat{h} \| \, |t|}{d} \right)^d \; e^{2 |t| \|  \hat{h} \|  d} \;,
\end{equation} 
which explicitly shows that  the dependence from the system size present 
in~(\ref{eq: Poly bound1}) is lost in favour of a dependence on the interaction strength $\|  \hat{h} \|$ similar to what we observed in Sec.~\ref{sec1new}.
In particular  for  small times the new inequality  mimics the polynomial behaviour of~(\ref{eq:MinNachBound}): as a matter of fact, in this regime, due to the presence of the  multiplicative
term $1/\sqrt{d}$, Eq.~(\ref{UUPP}) tends to be more strict than our previous bound (a result which is 
 not surprising as the derivation of  the present section takes full advantage of the linear topology of the network, while the analysis of Sec.~\ref{sec1new} holds true for a larger, less regular, class of possible scenarios). 
At large times on the contrary the new inequality 
is dominated by the exponential trend $e^{2 |t| \|  \hat{h} \|  d}$ which however tends to be overruled by 
the trivial bound (\ref{TRIVIAL}).

\subsection{A lower bound}\label{Sec.low} 
By properly handling  the identity~(\ref{eq: CBH expansion11}) it is also possible to derive a lower bound for $\| [\hat{A}(t),\hat{B}] \|$. 
Indeed  using the inequality
 $\|  \hat{O}_1+\hat{O}_2\|  \geq \| \hat{ O}_1\| -\| \hat{ O}_2\| $ we can  write 
\begin{eqnarray}\label{eq: Lower Bound1}
&&\|  [\hat{A}(t),\hat{B}] \|  =\Big\|  \sum\limits_{k=d}^\infty \frac{(it)^k}{k!} [[\hat{H},\hat{A}]_k,\hat{B} ]\Big\|  \\ \nonumber 
&&\geq   \frac{|t|^d}{d!}\|  [[\hat{H},\hat{A}]_d,\hat{B}]  \| -\Big\|  \sum\limits_{k=d+1}^\infty \frac{(it)^k}{k!} [[\hat{H},\hat{A}]_k,\hat{B} ] \Big\|  \;,
\end{eqnarray} 
(notice that the above bound is clearly trivial if  $[[\hat{H},\hat{A}]_d,\hat{B}]$ is the null operator: when this happens however we can replace it by substituting $d$ on it with the smallest $k>  d$ for which 
$[[\hat{H},\hat{A}]_k,\hat{B}]\neq 0$).
Now we observe that the last term appearing on the r.h.s. of the above expression can be
bounded by following the same derivation of the previous paragraphs, i.e. 
\begin{eqnarray}
&&\Big\|  \sum\limits_{k=d+1}^\infty \frac{(it)^k}{k!} [[\hat{H},\hat{A}]_k,\hat{B}] \Big\|   \nonumber \\
&&\quad\qquad \leq   2\|  \hat{A} \|  \|  \hat{B} \|  \sum_{k=d+1}^\infty n_k\frac{(2 |t| \|  \hat{h} \| )^k}{k!} \nonumber \\
&&\quad\qquad \leq 
2\|  \hat{A} \|  \|  \hat{B} \| \frac{\left( 2 |t| \|  \hat{h} \| \right)^d}{d!}\sum_{k=1}^\infty \frac{\left( 2 |t| \|  \hat{h} \|  d\right)^k}{k!}\nonumber \\ 
&&\quad \qquad =2\|  \hat{A} \|  \|  \hat{B} \| \frac{\left(2 |t| \|  \hat{h} \| \right)^d}{d!}(e^{2 |t| \|  \hat{h} \|  d}-1)  \nonumber\\
&&\quad\qquad \leq  2\|  \hat{A} \|  \|  \hat{B} \| {\left(\frac{2 e |t| \|  \hat{h} \|}{d} \right)^d}\frac{e^{2 |t| \|  \hat{h} \|  d}-1}{\sqrt{2 \pi d} } \;. \nonumber \end{eqnarray} 
Hence by replacing this into Eq.~(\ref{eq: Lower Bound1}) we obtain
\begin{eqnarray}\nonumber 
&&\|  [\hat{A}(t),\hat{B}] \|  \geq   \frac{ |t|^d}{d!}\|  [[\hat{H},\hat{A}]_d,\hat{B}]  \| 
 \\
&&\qquad  - 2\|  \hat{A} \|  \|  \hat{B} \| {\left(\frac{2 e |t| \|  \hat{h} \|}{d} \right)^d}\frac{e^{2 |t| \|  \hat{h} \|  d}-1}{\sqrt{2 \pi d} }
\nonumber \\
&&\qquad  \geq  \frac{2\|  \hat{A} \|  \|  \hat{B} \|}{  \sqrt{2 \pi d} } \nonumber 
\left(\frac{2 e  |t|  \| \hat{h} \|}{d} \right)^d \left( \Gamma_d - ({e^{2 |t| \|  \hat{h} \|  d}-1}) \right) \;, \\ 
\label{eq: Lower Bound1new}
\end{eqnarray}
where in the last passage we used  the upper bound for $d!$ that comes from Eq.~(\ref{STIRLING}) and introduced the dimensionless quantity 
\begin{eqnarray} \label{defgamma} 
\Gamma_d: =\sqrt{ \frac{\pi}{2 e^2}} \;  \frac{ \| [[\hat{H},\hat{A}]_d,\hat{B}]  \|}{ \|  \hat{A} \|  \|  \hat{B} \|  (2 \| \hat{h}\|)^d} \;,
\end{eqnarray} 
which can be shown to be strictly smaller than $1$ (see Sec.~\ref{sec. Counting comm}).

It's easy to verify that as long as $\Gamma_d$ is non-zero (i.e. as long as $[[\hat{H},\hat{A}]_d,\hat{B}]\neq 0$), 
there exists always a sufficiently small time  $\bar{t}$ such that $\forall\, 0<t<\bar{t}$
the r.h.s. of Eq.~(\ref{eq: Lower Bound1new}) is explicitly positive, implying that 
 we could have a finite amount of correlation at a time shorter than that required to light pulse to travel from $A$ to $B$ at speed $c$.  This apparent violation of causality is clearly a consequence of the approximations
 that lead to the effective spin Hamiltonian we are working on (the predictive power of the model
 being always restricted to time scales $t$ which  are larger than  
 $\frac{d(A,B)}{c}$). 
 More precisely, for sufficiently small value of $t$ (i.e.  for 
 $2 |t|\|  \hat{h} \|  d \ll 1$)  the negative contribution 
 on the r.h.s. of  Eq.~(\ref{eq: Lower Bound1new}) can be neglected and the bound predicts 
 the norm of $ [\hat{A}(t),\hat{B}]$ to grow polynomially as $t^d$, i.e.  
\begin{eqnarray}\label{eq: Lower Bound1new1}
 \| [\hat{A}(t),\hat{B}] \|  
&\gtrsim&   \frac{2\|  \hat{A} \|  \|  \hat{B} \|}{  \sqrt{2 \pi d} } 
\left(\frac{2 e  |t|  \| \hat{h} \|}{d} \right)^d  \Gamma_d  \;,
\end{eqnarray}
which  should be compared with 
\begin{eqnarray}\label{eq: upper Bound1new1}
\|  [\hat{A}(t),\hat{B}] \|  
&\lesssim&   \frac{2\|  \hat{A} \|  \|  \hat{B} \|}{  \sqrt{2 \pi d} } 
\left(\frac{2 e  |t|  \| \hat{h} \|}{d} \right)^d   \;,
\end{eqnarray}
that, for the same temporal regimes is instead predicted from the upper bound~(\ref{UUPP}). 

\subsection{Counting commutators}\label{sec. Counting comm}

\begin{figure}[t!]

\begin{tabular}{@{}c@{}}
   \includegraphics[width=\linewidth]{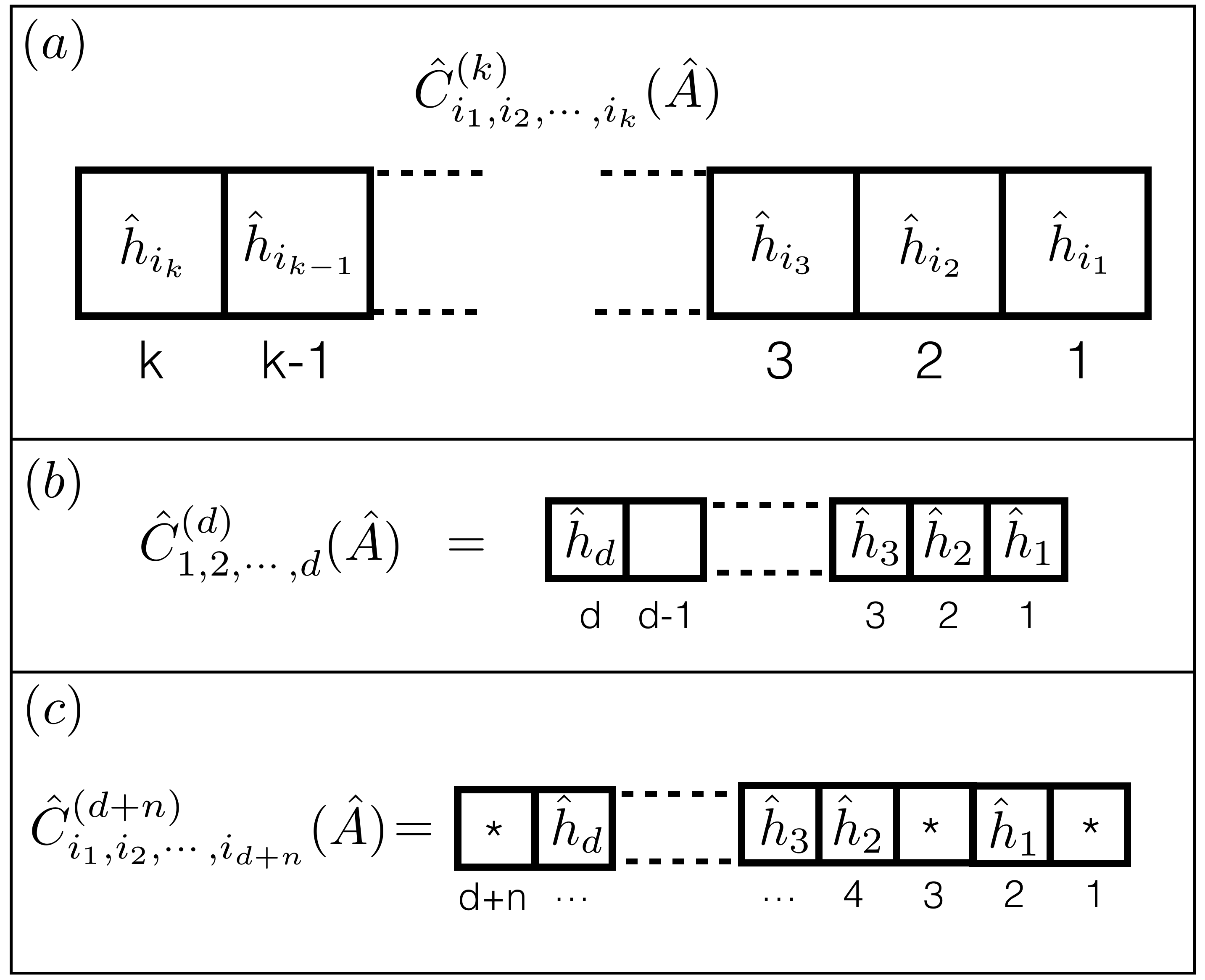} 
   \end{tabular}
  \caption{Panel (a): Pictorial representation of the  nested commutator $\hat{C}^{(k)}_{1,2 ,\cdots ,k} (\hat{A})$ as a set of boxes, each one fillable with a $\hat{h}_i$;
  Panel (b): representation of the 
  only nested commutator which for the case $k=d$  admits a non-zero value for the commutation with $\hat{B}$;
  Panel (c): case $k=d+n$ with $n\geq 1$. Here the boxes indicated with the asterisk can be filled  depending on their position, for instance here the box before $\hat{h}_1$ could contain only $\hat{h}_1$ while the one after $\hat{h}_d$ could contain any.}
  \label{fig:boxes}
\end{figure}

Here we report the explicit derivation of the 
 inequality~(\ref{BOUNDONNK}). The starting point of the analysis is 
the recursive identity 
\begin{eqnarray} \label{recursive} 
\hat{C}^{(k)}_{i_1,i_2 ,\cdots ,i_k} (\hat{A}) = [\hat{h}_{i_k} , \hat{C}^{(k-1)}_{i_1,i_2 ,\cdots ,i_{k-1}} (\hat{A})]\;,
\end{eqnarray} 
which links the expression for nested commutators~(\ref{Knested}) of order $k$ to those of order $k-1$. 
Remember now  that the operator $\hat{A}$ is located on the first site of the chain.  Accordingly, from 
Eq.~(\ref{COMM}) it follows that 
\begin{eqnarray} \hat{C}^{(1)}_{i} (\hat{A}) = [\hat{h}_{i} , \hat{A}]=0  \;, \qquad \forall i\geq 2\;, \end{eqnarray}  
i.e. the only possibly non-zero nested commutator of order 1 will be the operator $\hat{C}^{(1)}_{1} (\hat{A})=[\hat{h}_{1} , \hat{A}]$ which acts non trivially on the first and second spin. 
From this and the recursive identity~(\ref{recursive})  we can then derive the following identity for the nested commutator of order $k=2$, i.e. 
\begin{eqnarray} 
\hat{C}^{(2)}_{1,i_2} (\hat{A})&=&0\;,   \qquad \forall i_2 \geq 3 \;, \\ 
\hat{C}^{(2)}_{i_1,i_2} (\hat{A})&=&0\;,   \qquad \mbox{$\forall i_1 \geq 2$ {and}  $\forall i_2\geq 1$} \;, 
\end{eqnarray} 
 the only  terms  which can be possibly non-zero being now $\hat{C}^{(2)}_{1,1} (\hat{A})$
 and $\hat{C}^{(2)}_{1,2} (\hat{A})= [\hat{h}_{2} ,[ \hat{h}_{1},\hat{A}]]$, the first having support 
 on the first and second spin of the chain, the second instead being supported on the first, second, and third spin. Iterating the procedure it turns out  that for generic value of $k$, the operators  
$\hat{C}^{(k)}_{i_1,i_2 ,\cdots ,i_k} (\hat{A})$ which may be explicitly not null are those for which we have 
\begin{equation} \left\{ 
\begin{array}{l} 
i_1 = 1\;, \\ 
 i_j \leq \max\{ i_1, i_2, \cdots, i_{j-1}\} + 1\;,  \qquad \forall j\in\{ 2, \cdots, k\}\;,
\end{array} \right. \label{NONZERO} 
\end{equation} 
the rule being that passing from $\hat{C}^{(k-1)}_{i_1,i_2 ,\cdots ,i_{k-1}} (\hat{A})$
to $\hat{C}^{(k)}_{i_1,i_2 ,\cdots ,i_k} (\hat{A})$,
the new Hamiltonian element $\hat{h}_{i_k}$ entering (\ref{recursive}) 
has to be one of those already touched (except the first one $[\hat{h}_1,A]$) or one at distance at most 1 to the maximum position reached until there.
We also observe that among the  element $\hat{C}^{(k)}_{i_1,i_2 ,\cdots ,i_k} (\hat{A})$ which are not null, the one which have the largest support are those that have
the largest value of the indexes: indeed from (\ref{recursive}) it follows that 
the extra  commutator with $\hat{h}_{i_k}$  will create an operator whose support either
coincides with  the one 
of 
 $\hat{C}^{(k-1)}_{i_1,i_2 ,\cdots ,i_{k-1}} (\hat{A})$ 
 (this happens whenever  ${i_k}$ belongs to $\{ i_1,i_2 ,\cdots ,i_{k-1}\}$), or it is larger than the latter by one (this happens  instead  
for ${i_k} = \max\{ i_1, i_2, \cdots, i_{k-1}\} + 1$).
Accordingly among the nested commutators of order $k$ the one with the largest support is
\begin{eqnarray} 
\hat{C}^{(k)}_{1,2 ,\cdots ,k} (\hat{A}) = 
[\hat{h}_{k} , [\hat{h}_{{k-1}} , \cdots ,[\hat{h}_{2}, [\hat{h}_{1} ,\hat{A}]]\cdots]]\;,
\end{eqnarray} 
that in principle operates non trivially on all the first $k+1$ elements of the chain. 
Observe then that in order to get a non-zero contribution in (\ref{summation}) 
 we also need the succession $\hat{h}_{i}$ entering $\hat{C}^{(k)}_{i_1,i_2 ,\cdots ,i_k} (\hat{A})$ 
  to touch at least once the support of $\hat{B}$. This, together with the prescription just discussed, implies that at least once every element $\hat{h}_i$ between $A$ and $B$ has to appear, and the first appearance of each $\hat{h}_i$ has to happen after the first appearance of $\hat{h}_{i-1}$. 
In summary we can think each nested commutator of order $k$ as a numbered set of $k$ boxes fillable with elements $\hat{h}_i$ (see Fig.~\ref{fig:boxes} (a)) and, keeping in mind the rules just discussed, we want to count how many fillings give us non zero commutators.
Starting from $k=d$, we have only one possibility, i.e. the 
element $\hat{C}^{(d)}_{1,2 ,\cdots ,d} (\hat{A})$, see Fig.~\ref{fig:boxes} (b). This implies  
\begin{eqnarray}
[[\hat{H},\hat{A}]_d,\hat{B}] &=& [\hat{C}^{(d)}_{1,2 ,\cdots ,d} (\hat{A}),\hat{B}] \\ \nonumber 
&=& 
[ [\hat{h}_{d} , [\hat{h}_{{d-1}} , \cdots ,[\hat{h}_{2}, [\hat{h}_{1} ,\hat{A}]]\cdots]],\hat{B}]\;,
\end{eqnarray} 
and hence by sub-additivity of the norm, to
\begin{eqnarray}\label{eq:Gamma bound}
\| [[\hat{H},\hat{A}]_d,\hat{B}]\| \leq  2  
\|  \hat{A} \|  
\|  \hat{B} \| 
 (2 \| \hat{h}\|)^d\;,
\end{eqnarray} 
which leads to $\Gamma_d\leq \sqrt{ {2 \pi}/{e^2}}\simeq 0.923$ as anticipated in the paragraph below Eq.~(\ref{defgamma}).
 Consider next the case  $k=d+n$ with $n\geq 1$. In this event we must have at least  $d$ boxes filled with each $\hat{h}_i$ between $\hat{A}$ and $\hat{B}$. 
 Once we fix them, the content of the remaining $k=n-d$ boxes (indicated by an asterisk in panel (c) of  Fig.~\ref{fig:boxes})
 depends on their position in the sequence: if one of those is before the first $\hat{h}_1$ it will be forced to be $\hat{h}_1$, if it's before the first $\hat{h}_2$ it will be $\hat{h}_1$ or $\hat{h}_2$ and so on until the one before the first $\hat{h}_{d}$, which will be anyone among the $\hat{h}_i$.
  So in order to compute the number $n_k$ of non-zero terms entering (\ref{summation}) 
    we need to know in how many ways we can dispose the empty boxes in the sequence: since empty boxes (as well as the ones necessarily filled) are indistinguishable  there are $\binom{k}{n}=\binom{k}{d}$ ways. For each way we'd have to count possible fillings, but there's not a straightforward method to do it so we settle for an upper bound. The worst case is the one in which all empty boxes come after the first $\hat{h}_{d}$, so that we have $d^n$ fillings, accordingly we can bound $n_k$ with 
$\binom{k}{n}d^n =\binom{k}{d}d^{k-d}$ leading to 
 Eq.~(\ref{BOUNDONNK}).

As mentioned at the beginning of the section  a technique similar to the one reported here 
has been presented in the recent literature expressed in \cite{FinTemp2,Them}. These works also
results in a polynomial upper bound for the commutator, yet  it appears that the number of contributions entering in the parameter $n_k$ could be underestimated, and this underestimation is negligible only at orders $k\simeq d$ or, equivalently, at small times.  Specifically in \cite{Them}, which exploits intermediate results from \cite{ThemRef,ExistDynam}, the bound is obtained from the iteration of the inequality:
\begin{equation}
C_B(t,X)\leq C_B(0,X)+2\sum_{Z\in \partial X}\int_0^{|t|} \mathrm{d}s \, \, C_B(s,Z)\|  \hat{H}_Z \| ,
\end{equation}
where $C_B(t,X)=\|  [A(t),\hat{B}] \| $, $X$ is the support of $A$ and $\partial X$ is the surface of the set $X$. The iteration adopted in  \cite{Them} produces  an object that involve a summation of the form $\sum\limits_{Z\in \partial X}\,\sum\limits_{Z_1\in \partial Z}\,\sum\limits_{Z_2\in \partial Z_1}\cdots$. This selection however  underestimates the actual number of contributing terms. Indeed in the first order of iteration $Z\in \partial X$ takes account of all Hamiltonian elements non commutating with $\hat{A}$, but the next iteration needs to count all non commuting elements, given by $Z_1\in \partial Z$ \textit{and}  $Z\in \partial X$. So the generally correct statement, as in Ref.~\cite{ExistDynam}, would be $\sum\limits_{Z\cap X\neq \emptyset}\,\,\sum\limits_{Z_1\cap Z\neq \emptyset}\,\,\sum\limits_{Z_2\cap Z_1\neq \emptyset }\cdots$. 
The above discrepancy is particularly evident when 
focusing on the linear spin chain case we consider here. Taking account only of surface terms in the nested commutators in Eq.~(\ref{eq: Corr func all terms}),  among all the contributions which can be non-zero according to Eq.~(\ref{NONZERO}), 
we would have included only those with  $i_{j+1}=i_j+1$. This corrections are irrelevant at the first order in time in Eq.~(\ref{eq: Corr func all terms}) but lead to underestimations in successive orders. In \cite{Them} the discrepancy is mitigated at first orders by the fact that the number of paths of length $L$ considered is upper bounded by $N_1(L):=(2(2\delta-1))^L$ with $\delta$ dimensions of the graph. But again at higher orders this quantity is overcome by the actual numbers of potentially not null commutators (interestingly in the case of 2-D square lattice $N_1(L)$ could be found exactly, shrinking at the minimum the bound, see \cite{Guy}). Similarly is done in \cite{FinTemp2}, where, in the specific case of a 2-D square lattice, to estimate the number of paths of length $L$ a coordination number $C$ is used, which gives an upper bound $N_2(L):=(2C-1)^L$ that for higher orders is again an underestimation. To better visualize why this is the case, let's consider once more the chain configuration. Following rules of Eq.~(\ref{NONZERO}) 
 we understood that nested commutators
$\hat{C}^{(k)}_{i_1,i_2 ,\cdots ,i_k} (\hat{A})$  with repetitions of indexes. So with growing $k$ the number of possibilities for successive terms in the commutator grows itself: this is equivalent to a growing dimension $\delta^{(k)}$ or coordination number $C^{(k)}$. For instance we can study the multiplicity of the extensions of the first not null order $\hat{C}^{(d)}_{1,2 ,\cdots ,d} (\hat{A})$. Since the support of this commutator has covered all links between $A$ and $B$ we can choose among $d$ possibilities (not taking into account possible sites beyond $B$ and before $A$, depending on the geometry of the chain we choose), we'll have then $d^{L-d}$ possibilities at the $L$-th order: for suitable $d$ and $n$ we shall have $d^{L-d}>N_1(L),N_2(L)$. This multiplicity is relative to a single initial path, so we do not even need to count also the different possible initial paths one can construct with $d+l$ steps s.t. $d+l<L$. 

In summary, the polynomial behaviour found previously in the literature is solid at the first order but could not be at higher orders.

\section{Simulation for a Heisenberg XY chain} \label{Sec:Simulation}
\begin{figure}[t!]
  \includegraphics[width=\linewidth]{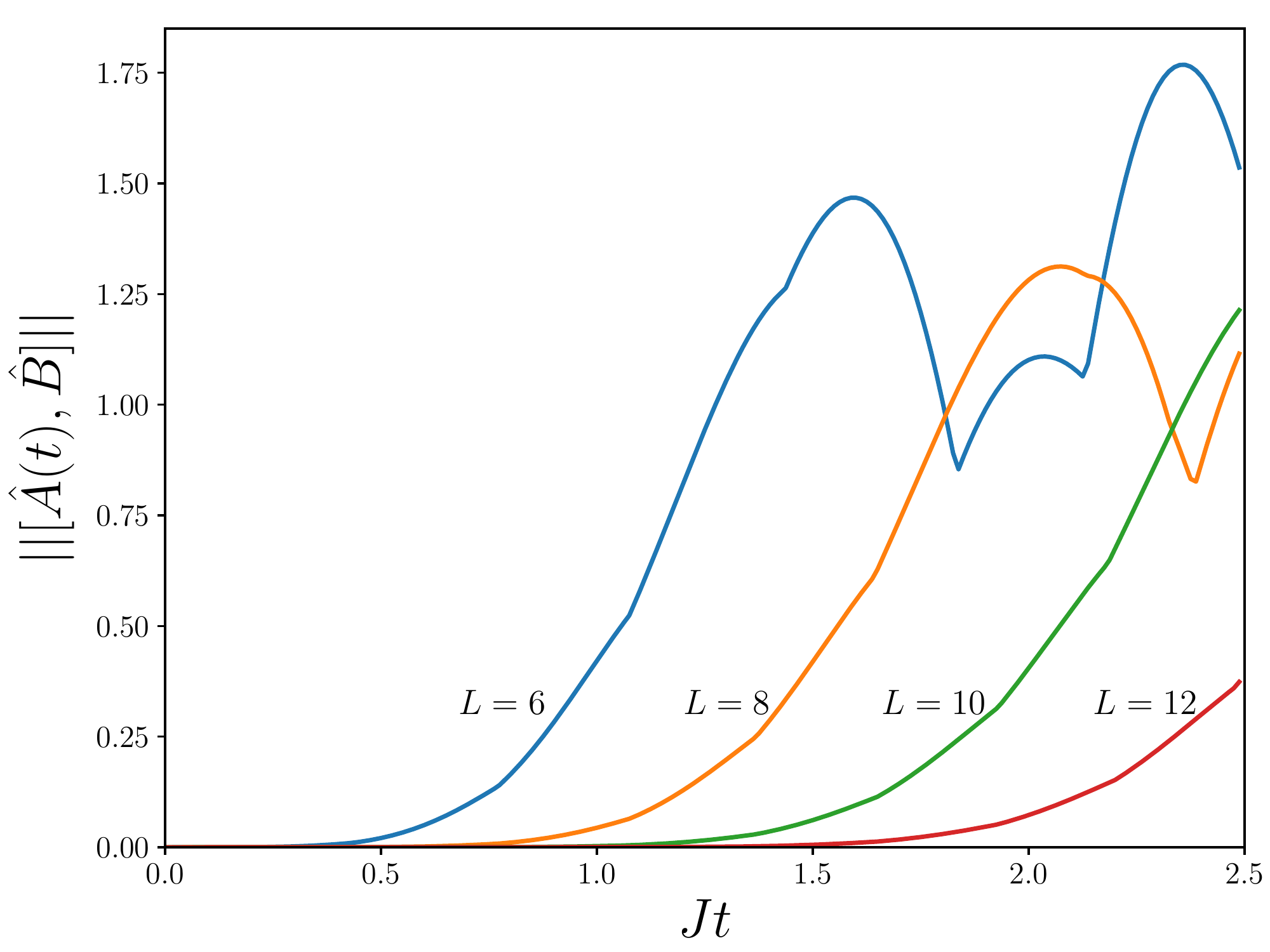}
   \vspace{-20pt}
  \caption{(Color online): Simulation of $\| [\hat{A}(t),\hat{B}]\|$ for different chain lengths $L$ for the  Heisenberg XY linear spin-chain. }
   \label{fig:plot_sim}
\end{figure}

Here we  test the validity of our results presented in the previous section for a reasonably simple system such as a uniformly coupled, next-neighbour Heisenberg  XY chain composed by $L$ spin-$1/2$, described by the following Hamiltonian:
\begin{equation}
\hat{H}=J\sum_{i=0}^{L-1}\hat{\sigma}_i^x \hat{\sigma}_{i+1}^x+\hat{\sigma}_i^y \hat{\sigma}_{i+1}^y\;. \label{DEFHAMH} 
\end{equation}
As local operators $\hat{A}$ and $\hat{B}$ we adopt two $\hat{\sigma}^z$ operators, acting respectively on the first and last spin of the chain, so that $\|\hat{A}\|=\|\hat{B} \|=1$. Employing QuTiP~\cite{Qutip1,Qutip2} we perform the numerical evaluation for $\|[\hat{A}(t),\hat{B}]\|$ varying the length of the chain $L$. 
 (Fig.~\ref{fig:plot_sim}).

\begin{figure}[t!]
  \includegraphics[width=\linewidth]{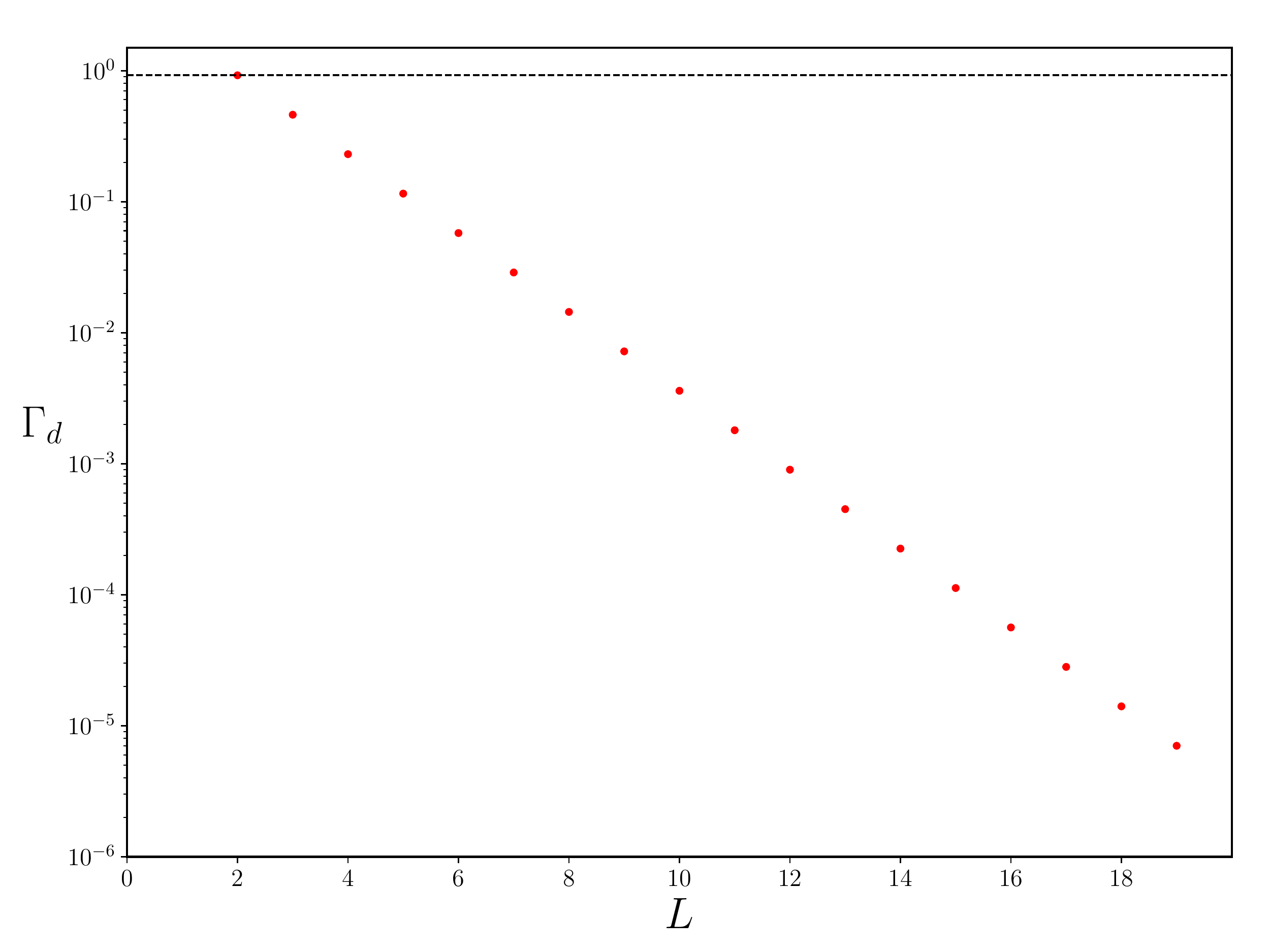}
  \vspace{-20pt}
  \caption{(Color online): Plot of the value of  $\Gamma_d$ defined in  Eq.~(\ref{defgamma}) for different values  of the chain length $L$,  $d$ being fixed equal to $L-1$. Notice that all values are below 
   $\sqrt{ {2 \pi}/{e^2}}$ (dashed line) which is provably the largest value this parameter can achieve.}
  \label{fig:plot_Gamma}
\end{figure}

\begin{figure}[b!]
  \includegraphics[width=\linewidth]{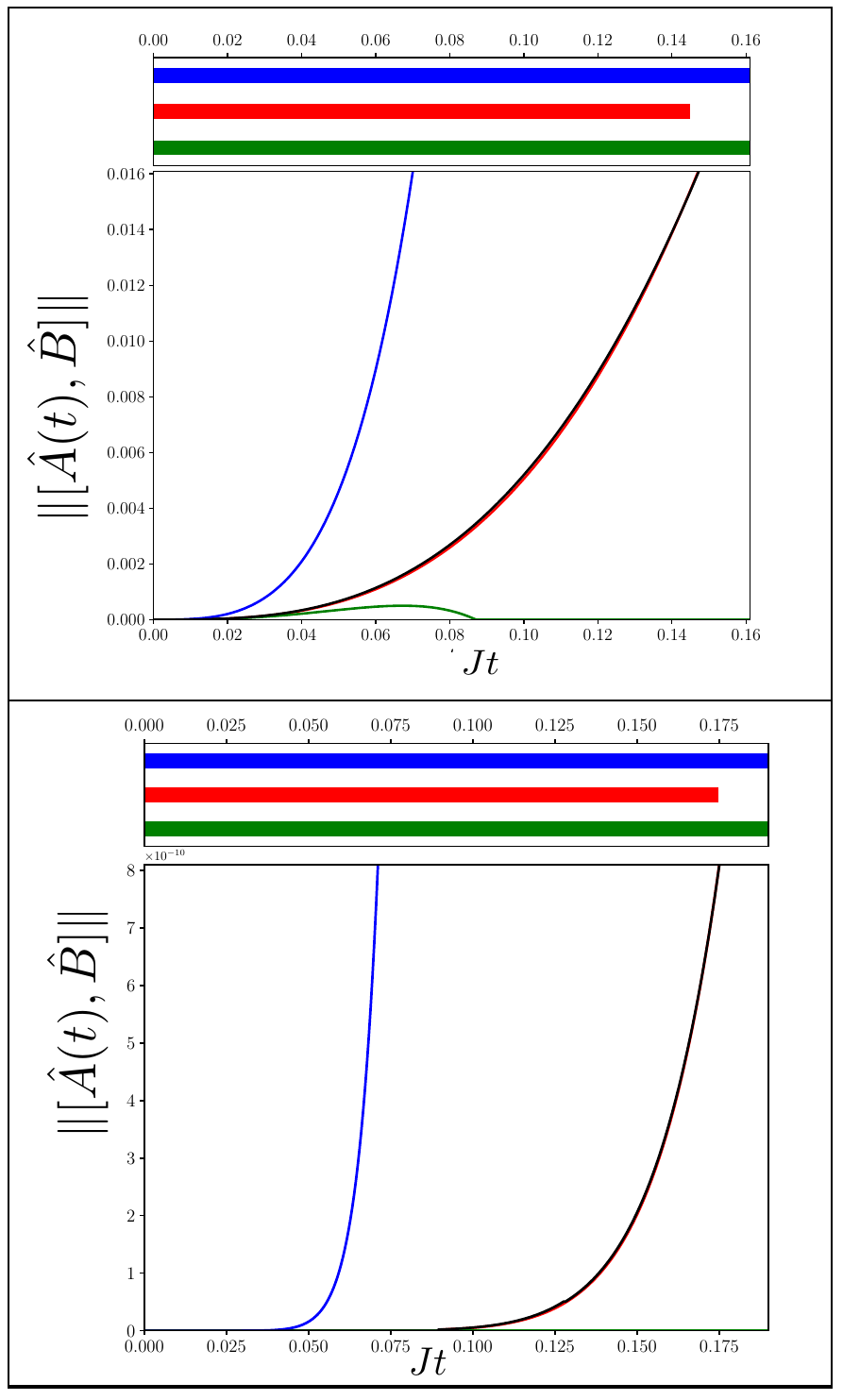}
  \caption{(Color online): Simulation and bounds of the function $\|[\hat{A}(t),\hat{B}]\|$
  for a $L=4$ spin-chain (upper panel) and for a $L=10$ spin-chain (lower panel). 
    The plot shows upper bound~(\ref{UUPP}) (blue curve) lower bound~(\ref{eq: Lower Bound1new}) (green curve), simplified lower bound (\ref{eq: Lower Bound1new1}) (red) and numerical simulation (black). 
  The coloured bars above the plots outline the time domain in which the each bound (identified by the same colors) results to be valid.  As expected the simplified bound stands only for sufficiently small times.
  In all cases the simulation  and the simplified lower bound  are comparable in magnitude so that their curves are hardly distinguishable.
    In the case of $L=10$  the complete lower bound ~(\ref{eq: Lower Bound1new}) (green) is considerably small, hence not visible.  }
  \label{fig:L=4}
\end{figure}
We are interested in the comparison between these  results with the expressions obtained for the upper bound (\ref{UUPP}), the lower bound (\ref{eq: Lower Bound1new}), and the simplified lower bound at short times (\ref{eq: Lower Bound1new1}). The time domain in which the simplified lower bound stands depends also on the value of the parameter $\Gamma_d$ specified in Eq.~(\ref{defgamma}), which we understood to be $\leq \sqrt{2\pi /e^2}$ but which we need reasonably large in order to produce a detectable bound in the numerical evaluation. In Fig.~\ref{fig:plot_Gamma} values of $\Gamma_d$ for different chain lengths $L$ (s.t. $d=L-1$) are reported. The magnitude of $\Gamma_d$ exhibits an exponential decrease with the size of the chain $L$. 
The results of our simulations are presented in Fig.~\ref{fig:L=4} for the cases $L=4$ and $L=10$. 
The upper bound (\ref{UUPP}), as well as the lower bound (\ref{eq: Lower Bound1new}) should result to be universal, i.e. to hold for every $t$, although being the latter trivial at large times. This condition is satisfied for every $L$ at every $t$ analysed (we performed the simulation for $2\leq L \leq 12$). For what concerns the simplified lower bound (\ref{eq: Lower Bound1new1}), we would expect its validity to be guaranteed only for sufficiently small $t$ and as a matter of fact we find the time domain of validity to be limited at relatively small times (see e.g. the histograms in Fig.~\ref{fig:L=4}).

\section{Conclusions} \label{Sec:conc} 
The study 
of the L-R inequality  we have presented here shows that for a large class of spin-network models characterized by couplings that are of finite range, 
 the correlation function $\|[\hat{A}(t),\hat{B}]\|$ can be more tightly bounded by a new constraining function  that exhibits a polynomial dependence with respect to time,
and which, for sufficiently large distances,  allows for a precise definition of a maximum speed of the signal propagation, see Eq.~(\ref{SPEED}).
Our approach does not rely on often complicated   graph-counting arguments, instead is based on an analytical optimization of the original inequality~\cite{LR} 
with respect to all free parameters of the model (specifically the $\lambda$ parameter defining via Eq.~(\ref{DEFPHILambda}) the convergence of the Hamiltonian couplings at large
distances).  
Yet, in the special case of
linear spin-chain, we do adopt a graph-counting  argument  to present an alternative derivation of our result and to show that a similar reasoning can be used
to also construct non-trivial lower bounds for  $\|[\hat{A}(t),\hat{B}]\|$ when the two sites are located at the opposite ends of the chain. 
Possible generalizations of the present approach can be foreseen by including a refined evaluation of the dependence upon $\lambda$ of Eq.~(\ref{DEFPHILambda}), that goes
beyond the one  we adopted in Eq.~(\ref{eq: Max diameter}).

We point out that during the preparation of this manuscript the same result presented in Eq.~(\ref{eq: upper Bound1new1}) for a chain appeared in Ref.~\cite{Lucas}.

The Authors would like to thank R. Fazio and B. Nachtergaele for their comments and suggestions.


\begin{thebibliography}{1}

\bibitem{BOSE1} S. Bose, Contemp. Phys. 48, 13 (2007).


\bibitem{LR} E. Lieb and  D. Robinson,
 Commun. Math. Phys. {\bf 28}, 251-257, (1972).
 
 \bibitem{REVIEW} 
B. Nachtergaele, R. Sims, and A. Young, 	arXiv:1810.02428 [math-ph].

\bibitem{LSM Theo} M. B. Hastings, Phys. Rev. B {\bf 69}, 104431, (2004). % \href{https://arxiv.org/abs/cond-mat/0305505}{arXiv:0305505}.

\bibitem{Clust Theo} B. Nachtergaele and R. Sims, Commun. Math. Phys. {\bf 265}, 119  (2006). %	\href{https://arxiv.org/abs/math-ph/0506030}{arXiv:0506030}.

\bibitem{ExistDynam} B. Nachtergaele, Y. Ogata, R. Sims, J. Stat. Phys., 124: 1, (2006). %\href{https://arxiv.org/abs/math-ph/0603064}{arXiv:0603064}.

\bibitem{exponential1} M. B. Hastings and T. Koma, Commun. Math. Phys.  {\bf 265}, 781 (2006). %\href{https://arxiv.org/abs/math-ph/0507008}{arXiv:0507008}.

\bibitem{AreaLaw} M. B. Hastings, J. Stat. Mech. {\bf 8}, P08024, (2007).% \href{https://arxiv.org/abs/0705.2024}{	arXiv:0705.2024}.

\bibitem{TopQOrder} S. Bravyi, M. B. Hastings, and S. Michalakis, J. Math.  Phys. {\bf 51}, 9 (2010).         %   \href{https://arxiv.org/abs/1001.0344}{arXiv:1001.0344}.

\bibitem{BRAV} S. Bravyi,  M. B. Hastings, and F. Verstraete.  Phys. Rev. Lett. {\bf 97}, 050401  (2006).
\bibitem{SUPER} J. Eisert and D. Gross,  Phys. Rev. Lett. {\bf 102}, 240501 (2009).
\bibitem{EISERT} J. Eisert, {\it et al.} Phys. Rev. Lett. {\bf 111}, 26401 (2013).
\bibitem{PRA} J. M. Epstein and K. B. Whaley, Phys. Rev. A {\bf 95}, 042314 (2017). 




\bibitem{LongRange} P. Hauke and L. Tagliacozzo,
Phys. Rev. Lett. {\bf 111}, 207202 (2013). \href{https://arxiv.org/abs/1304.7725}.

\bibitem{LongRange1} J. Eisert, M. van den Worm,  S. R. Manmana, and M. Kastner
Phys. Rev. Lett. {\bf 111}, 260401 (2013). %     \href{https://arxiv.org/abs/1309.2308}{arXiv:1309.2308}.

\bibitem{LongRange2} Z.-X.Gong, M. Foss-Feig, S. Michalakis, and A. V. Gorshkov, 
Phys. Rev. Lett. {\bf 113} , 030602 (2014).    
%\href{https://arxiv.org/abs/1401.6174}{arXiv:1401.6174}.

\bibitem{LongRange3} M. Foss-Feig, Z.-X. Gong, C. W. Clark, and A. V. Gorshkov
Phys. Rev. Lett. {\bf 114}, 157201 (2015). 
% \href{https://arxiv.org/abs/1410.3466}{arXiv:1410.3466}.

\bibitem{LongRange4} T. Matsuta, T. Koma, S. Nakamura, Ann. Henri Poincar\'e {\bf 18}, 519
(2017). 
%\href{https://arxiv.org/abs/1604.05809}{arXiv:1604.05809}.

\bibitem{Burrell} C. Burrell and T. Osborne, Phys. Rev. Lett. {\bf 99}, 167201, (2007). % \href{https://arxiv.org/abs/quant-ph/0703209}{	arXiv:0703209}.

\bibitem{Burrell2} C. K. Burrell, J. Eisert, and T. J. Osborne,
Phys. Rev. A {\bf 80}, 052319 (2009). %  \href{https://arxiv.org/abs/0809.4833}{arXiv:0809.4833}.

\bibitem{FinTemp} M. B. Hastings,  Phys. Rev. Lett. {\bf 93}, 126402 (2004).  % \href{https://arxiv.org/abs/cond-mat/0406348}{arXiv:0406348}.

\bibitem{FinTemp1} S. Hernandez-Santana, C. Gogolin, J. I. Cirac, and A. Ac\`{\i}n,
Phys. Rev. Lett. {\bf 119}, 110601 (2017).
 %\href{https://arxiv.org/abs/1702.00371}{arXiv:1702.00371}.

\bibitem{FinTemp2} Z. Huang and X.-K. Guo,  Phys. Rev. E {\bf 97}, 062131 (2018).
% \href{https://arxiv.org/abs/1711.06977}{arXiv:1711.06977}.

\bibitem{Scram}  N. Lashkari, D. Stanford, M. Hastings, T. Osborne, and P. Hayden, J. High Energ. Phys. {\bf 04}, 022 (2013).

\bibitem{Scram1} D. A. Roberts and B. Swingle
Phys. Rev. Lett. {\bf 117}, 091602 (2016).

\bibitem{Them} K. Them, Phys. Rev. A {\bf 89}, 022126 (2014). 
%\href{https://arxiv.org/abs/1308.2882}{arXiv:1308.2882}.



\bibitem{Bratteli} O. Bratteli and D. W. Robinson, 
 {\it Operator Algebras and Quantum Statistical Mechanics}  (Springer Verlag, 1997).
 
 
 \bibitem{NACHTER1} B. Nachtergaele and R. Sims, in: {\it New Trends in Mathematical Physics}. Selected contributions of the XV-th International Congress on Mathematical Physics, 591--614 (Sidoravicius, Vladas (Ed.), Springer Verlag, 2009); 	arXiv:0712.3318 [math-ph].
 
 
 \bibitem{CRAMER} 
M. Cramer, A. Serafini, and J. Eisert, in {\it Quantum information and many body quantum systems}, 
Eds. M. Ericsson, S. Montangero, Pisa: Edizioni della Normale, pp 51-72, 2008 (Publications of the Scuola Normale Superiore. CRM Series, 8).
 
 
 \bibitem{ThemRef} B. Nachtergaele, H. Raz, B. Schlein, and R. Sims, Commun. Math. Phys. {\bf 286}, 1073 (2009). 
% \href{https://arxiv.org/abs/0712.3820}{arXiv:0712.3820}.
 

\bibitem{Guy} R. K. Guy, C. Krattenthaler, and B. E. Sagan,
 %Lattice paths, reflections, and dimension-changing bijections, 
 Ars. Combin. {\bf 34} 3, (1992).


\bibitem{Qutip1} J. R. Johansson, P. D. Nation, and F. Nori,
% "QuTiP 2: A Python framework for the dynamics of open quantum systems.",
 Comp. Phys. Comm. {\bf 184}, 1234 (2013).
 
\bibitem{Qutip2} J. R. Johansson, P. D. Nation, and F. Nori,
% "QuTiP 2: A Python framework for the dynamics of open quantum systems.",
Comp. Phys. Comm. {\bf 183}, 1760-1772 (2012).

\bibitem{Lucas} Chi-Fang Chen, A. Lucas, arXiv preprint  	arXiv:1905.03682 [math-ph], (2019).




\end{thebibliography}
\end{document}